%% file: csi_covariance_estimation_globecom_v1.3.tex
\documentclass[conference]{IEEEtran}
\ifCLASSINFOpdf
 \usepackage[pdftex]{graphicx}
\usepackage{pgfplots}
  \graphicspath{{fig/}{../jpeg/}}
\else
\fi
\usepackage[cmex10]{amsmath}
\usepackage{amsfonts}
\usepackage{setspace}
\usepackage[font=footnotesize]{subfig}

\begin{document}
%
\title{Channel Covariance Estimation in Massive MIMO Frequency Division Duplex Systems}

\author{\IEEEauthorblockN{Alexis Decurninge, Maxime Guillaud}
\IEEEauthorblockA{Huawei Technologies, Co. Ltd.\\
France Research Center\\
Mathematical and Algorithmic Sciences Lab  \\
\texttt{\small \{alexis.decurninge,maxime.guillaud\}@huawei.com } } 
\and
\IEEEauthorblockN{Dirk Slock}
\IEEEauthorblockA{EURECOM\\
Mobile Communications Department\\
\texttt{\small dirk.slock@eurecom.fr  } 
}}

%


\maketitle

\begin{abstract}
Channel covariance is emerging as a critical ingredient of the acquisition of instantaneous channel state information (CSI) in multi-user Massive MIMO systems operating in frequency division duplex (FDD) mode. In this context, channel reciprocity does not hold, and it is generally expected that covariance information about the downlink channel must be estimated and fed back by the user equipment (UE). As an alternative CSI acquisition technique, we propose to infer the downlink covariance based on the observed uplink covariance. This inference process relies on a dictionary of uplink/downlink covariance matrices, and on interpolation in the corresponding Riemannian space; once the dictionary is known, the estimation does not rely on any form of feedback from the UE. In this article, we present several variants of the interpolation method, and benchmark them through simulations.
\end{abstract}


%
\IEEEpeerreviewmaketitle

\makeatletter
\def\input@path{{fig/}{./}}
\makeatother

\def \R {\mathbf R}
\def \W {\mathbf W}
\def \H {\mathbf H}
\def \X {\mathbf X}
\def \Y {\mathbf Y}
\def \V {\mathbf V}
\def \M {\mathbf M}
\def \U {\mathbf U}
\def \h {\mathbf h}
\def \w {\mathbf w}
\def \h {\mathbf h}
\def \h {\mathbf h}
\def \RUL {\R^{\mathrm{UL}}}
\def \Ex {\mathrm{E}}


\section{Introduction}

Massive MIMO \cite{Marzetta_massive_MIMO} is a potential technology to provide the breakthroughs expected from fifth-generation (5G) wireless communication systems in terms of data rates and number of supported devices. Accurate and up-to-date channel state information (CSI) is a critical parameter in the operation of Massive MIMO \cite{Choi_Love_Bidigare_DL_training_FDD_MMIMO_JSTSP2014}. 
In the case of time-division duplex (TDD) operation, CSI can be obtained efficiently via channel reciprocity. For FDD systems, on the other hand, obtaining CSI requires over-the-air feedback\footnote{Note that direct exploitation of instantaneous channel reciprocity in FDD has been considered in \cite{Hochwald_Marzetta_UL_DL_FDD_reciprocity_sp2001} and \cite{Han_Ni_Du_reciprocity_FDD_chinacom2010}. However, these methods rely critically on assumptions about array geometry and orientation that make them generally impractical.}; in the Massive MIMO case, because of the dimensions of the involved channels, the required amount of downlink reference symbols and feedback have the potential to significantly reduce the overall spectral efficiency of the system.

Several authors \cite{yin12,adhikary13} have proposed to make use of second-order statistical information, and more precisely BTS-side channel covariance information, to optimize CSI acquisition. It has been shown that the amount of traning dedicated to instantaneous CSI estimation can be significantly reduced, when prior covariance information is available. Although these works do not cover the question of how the covariance information is obtained, covariance feedback (and quantization) has been extensively studied \cite{Krishnamachari_PhD_thesis}; however, continuous feedback of covariance information from the UE to the BTS reduces the fraction the uplink capacity actually available to transmit data. 
It is the object of the present article to propose a technique to estimate the covariance information on the downlink channel of a Massive MIMO system, based on the observed covariance of the uplink channel, i.e. without requiring continuous covariance feedback.
An approach to solve this problem was suggested in \cite{jordan09}, based on resampling the covariance for a different wavelength using cubic splines; however this method is only applicable to uniform linear arrays (ULA).
Conversely, the scheme proposed in this article is based on a dictionary of pairs of known uplink/downlink covariance matrices.  When a new uplink covariance matrix is observed, the corresponding downlink covariance is estimated through interpolation over a Riemannian space, using the elements of the dictionary. The advantages of the proposed scheme over state of the art techniques are:
\begin{itemize}
\item unlike \cite{jordan09}, no assumptions on the array geometry are required, and it is not required that the same set of BTS antennas are used for the uplink and downlink transmission;
\item it does not rely on feedback on the uplink, apart from an initial training phase;
\item it achieves a better estimation accuracy than the competing approaches (see Section~\ref{section_performance}).
\end{itemize}

In Section~\ref{section_model}, we introduce the system model. Several variants of the proposed interpolation method are discussed in Section~\ref{section_interpolation}.
The performance of the proposed scheme is benchmarked in Section~\ref{section_performance}, and its algorithmic complexity is discussed in Section~\ref{sec_complexity}.

Notations: $\Ex\{\cdot\}$ denotes the expectation, $(\cdot)^T$ and $(\cdot)^H$ respectively denote the transpose and the Hermitian transpose, and $\|.\|_F$  is the Frobenius norm. The set of complex Hermitian positive definite matrices of dimension $N\times N$ is denoted $\mathcal{S}_N^{++}(\mathbb{C})$.\\

\section{System Description and Model}
\label{section_model}

Let us consider the MIMO channel between a base station (BTS)  and a single-antenna user equipment (UE).
We assume that during the downlink transmission phase, the base station uses $N_T$ antennas to transmit, while during the uplink phase, the base station uses $N_R$ antennas to receive signals from the UE. Note that it is usually assumed that the same BTS antennas are used for the uplink and the downlink phases -- and indeed this can be construed as a special case of the considered setting; however the more general case considered in this article is of practical importance.
Let $\h$ denote the $N_T$-dimensional vector containing the channel coefficients between the BTS antennas and the single UE antenna, while $\h^{\mathrm{UL}}$ denotes the $N_R$-dimensional vector of uplink channel coefficients.

Using these notations, the scalar signal $y_{UE}$ received at the UE during downlink transmission can be written (omitting noise)
\begin{equation}
y_{\mathrm{UE}}=\h^T {\bf x}_{\mathrm{BTS}}
\end{equation}
where ${\bf x}_{BTS}$ is the $N_T$-dimensional vector of signals transmitted by the BTS, while during uplink transmission, the received $N_R$-dimensional signal at the BTS is 
\begin{equation}
{\bf y}_{\mathrm{BTS}}=\h^{\mathrm{UL}} x_{\mathrm{UE}}
\end{equation}
where $x_{\mathrm{UE}}$ denotes the scalar signal transmitted by the single-antenna UE.

Considering CSI estimation on the downlink, if no prior information about the channel statistics is available, a reference sequence of length at least $N_T$ is required to estimate $\h$. For a channel coherence time T, the fraction of channel use dedicated to data is upper bounded by $(T-N_T)/T$ which is problematic for large values of $N_T$ (massive MIMO regime).
In \cite{adhikary13}, it is considered how the knowledge of second-order channel statistics can help with CSI acquisition. Let us further assume that the coefficients of $\h$ are correlated, and that it can be written as
\begin{equation}
\h=\R^{1/2} \w 
\end{equation}
where
\begin{itemize}
\item $\R$ is the $N_T\times N_T$ BTS-side covariance matrix of the downlink channel (which is typically assumed constant for a given location of the UE)
\item $\w$ is an $N_T$-dimensional vector with independent, unit variance coefficients capturing the fast fading.
\end{itemize}
If $\R$ is known and rank limited (equal to $N_{RL}$), it is shown in \cite{TSP2004KotechaSayeed,adhikary13} that the length of the training sequence can  be reduced from $N_T$ to $N_{RL}$, thereby reducing the amount of pilot symbols required to estimate $\h$.\\

Similar to the downlink case where $\R = \Ex\left[\h \h^H \right]$, we let $\RUL$ denote the BTS-side correlation matrix of the uplink channel, i.e. $\RUL = \Ex\left[\h^{\mathrm{UL}} {\h^{\mathrm{UL}}}^H \right]$, of dimension  $N_R\times N_R$.



\section{Downlink Covariance Estimation Through Interpolation}
\label{section_interpolation}

We now describe the proposed downlink covariance estimation technique which is the object of this paper.

\subsection{General idea}

\begin{figure}
  \centering
   \subfloat[Learning phase (feedback performed for a set of UEs or UE positions).]{ \label{phase1}\input{learning_phase.tikz}}
    \hspace{1cm}   \subfloat[Exploitation phase (covariance CSI interpolated from stored dictionary).]{ \label{phase2}\input{exploitation_phase.tikz}}
 \caption{Illustration of the two phases.}
 \label{fig_phases}
\end{figure}
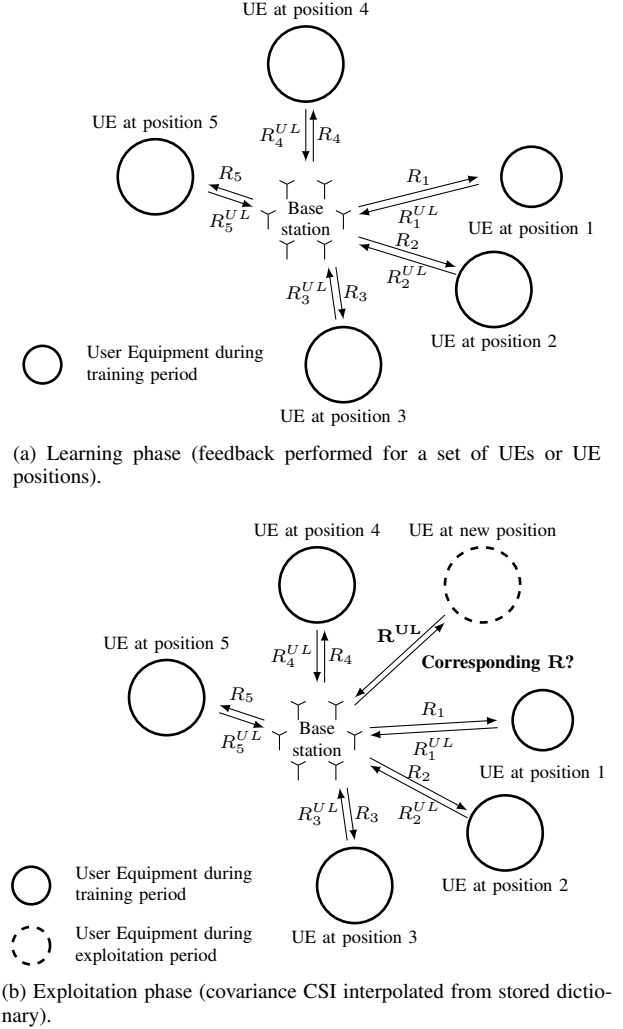

The method is based on a dictionary of covariance matrix pairs; let us denote $\R_i$ and $\R_i^{UL}$ the respective downlink and uplink covariance matrices corresponding to the $i$-th element of the dictionary (the index $i$ can refer to an instant of measurement or a position of the UE; see Figure \ref{fig_phases}). A general mathematical relation between $\R_i$ and $\R_i^{UL}$ does not exist, but since they originate from the same physical environment, we expect the same pair of uplink/downlink covariance matrices to occur again if the same or another UE happens to be in the same geometrical position at a later time. 
When a new uplink covariance matrix $\R^{UL}$ is observed on the uplink, we propose to estimate the corresponding downlink covariance $\R$ based on  $\R^{UL}$ and on the dictionary of known uplink/downlink covariances. 
Therefore, our scheme is divided in two phases:
\begin{itemize}
\item \textbf{A training phase}: at least one UE feeds back $\R_i$ for one or many positions such that we store a dictionary of matching pairs ($\R_i$, $\R_i^{UL}$) (Figure \ref{phase1}). The uplink and downlink covariances are assumed to have been measured for the same physical location of a given UE.
\item \textbf{An exploitation phase}: the BTS observes $\R^{UL}$ and interpolates $\R$ based on the stored dictionary (Figure \ref{phase2}). Thus, no feedback is involved in this phase.
\end{itemize}
The interpolation of the exploitation phase is performed in the space of Hermitian positive definite matrices viewed as a Riemannian space.

\subsection{Interpolation from a fixed dictionary of covariances}
\label{sec_interp}

During the exploitation phase, we assume that a dictionary of K downlink/uplink covariance matrix pairs $(\R_1, \R_1^{UL}),..., (\R_K, \R_K^{UL})$ is available thanks to the training mechanism already described.

\subsubsection{Estimator as a weighted barycenter}
For any uplink covariance $\R^{UL}$ observed during the exploitation phase, we deduce the corresponding $\R$ based on the dictionary by interpolating in the space of Hermitian positive definite matrices. We propose to define the interpolated value of the downlink covariance corresponding to $\R^{UL}$ as the barycenter of $\R_1,...,\R_K$ with weights $w_1,...,w_K$ defined by 
\begin{equation}
\hat{\R} = \arg\min_{\Y\in\mathcal{S}_{N_T}^{++}(\mathbb{C})} \sum_{i=1}^K w_id(\R_i,\Y)^2
\label{eq_barycenter}
\end{equation}
with $d(\cdot,\cdot)$ a distance defined in the space of positive definite matrices chosen amongst the distances presented in Table~\ref{figmetric}. The weights will be chosen based on the relative position of $\R^{UL}$ to the uplink covariance matrices from the dictionary (see Section~\ref{sec_weights} below).\\

\begin{table}
\centering
\doublespacing
\footnotesize
\begin{tabular}{|c|c|c|}
\hline
Metric & Distance $d(\X_1,\X_2)$ & Exponential map $ \exp_\X(\V)$\\
\hline
Euclidean&	$\|\X_1-\X_2\|_F$ & $ \X+\V$ \\
log-Euclidean &	$\|\log(\X_1)-\log(\X_2)\|_F$ & $ \exp(\log(\X)+\V)$\\
affine invariant  &	 $\|\log(\X_1^{\frac{1}{2}}\X_2^{-1}\X_1^{\frac{1}{2}})\|_F$ &	$\X^{\frac{1}{2}}\exp(\X^{-\frac{1}{2}}\V\X^{-\frac{1}{2}})\X^{\frac{1}{2}}$\\
\hline
\end{tabular}
\caption{Possible choices of distances and exponential maps for the Riemannian spaces $\mathcal{S}_{N_R}^{++}(\mathbb{C})$ and $\mathcal{S}_{N_T}^{++}(\mathbb{C})$. The operators $\log$ and $\exp$ stands for the matrix logarithm and exponential, respectively.}
\label{figmetric}
\end{table}

\subsubsection{Computation of the barycenter}

The minimization problem given by eq. (\ref{eq_barycenter}) should be solved by using the Riemannian metric associated to $d(\cdot,\cdot)$. Let us briefly introduce a basic tool of Riemannian optimization: the Riemannian exponential map $\V\mapsto \exp_\X(\V)$ is an operator indexed by $\X\in \mathcal{S}_N^{++}(\mathbb{C})$ mapping the tangent space at $\X$ (i.e. the space of gradient vectors originating at $\X$) onto an element of $\mathcal{S}_N^{++}(\mathbb{C})$, which is the result of the geodesic shooting from $\X$ with speed $\V$; see for example \cite{docarmo92} for more details on Riemannian tools.\\

Closed-form expression for the solution of \eqref{eq_barycenter} can be obtained when considering the Euclidean or log-Euclidean metric; the Euclidean metric yields
\begin{equation}
\hat{\R}_E = \frac{1}{K} \sum_{i=1}^K w_i\R_i,
\end{equation}
while the log-Euclidean metric yields
\begin{equation}
\hat{\R}_{LE} = \exp\left( \frac{1}{K} \sum_{i=1}^K w_i\log(\R_i)\right).
\end{equation}
As for the affine invariant metric, the barycenter has to be computed numerically through a gradient descent algorithm, as follows:
 we initialize the gradient descent with the log-Euclidean barycenter $\X_0=\hat{\R}_{LE}$. We then update iteratively $\X_t$ until convergence with (see for example \cite{pennec06b})
\begin{equation}
\X_{t+1} = \X_t^{\frac{1}{2}}\exp\left(\frac{1}{K}\sum_{i=1}^K w_i\log\left(\X_t^{-\frac{1}{2}}\R_i\X_t^{-\frac{1}{2}}\right)\right)\X_t^{\frac{1}{2}}.
\end{equation}

\subsubsection{Choice of the interpolation weights}
\label{sec_weights}

Obviously, the performance of the estimator depends critically on the choice of the weights in \eqref{eq_barycenter}. We consider three choices for the interpolation weights:
\begin{itemize}
\item \textbf{Nearest neighbor method} We let
\begin{equation}
w_i=\left\{\begin{array}{ll}1\text{   if  } i=\arg \min_p d(\R_p^{UL},\R^{UL}),\\
0 \text{    otherwise.}
\end{array}\right.
\end{equation}
In that case, the solution of the barycenter problem given by eq. (\ref{eq_barycenter}) is trivial because there is almost surely only one $w_i$ that is nonzero.

\item \textbf{Mirror-interpolation method}\\
The mirror-interpolation method consists in mirroring the interpolation in the space of uplink covariance matrices into the space of downlink matrices. Intuitively, it assumes that the structure of the dictionary of uplink covariance matrices is similar to the structure of the dictionary of downlink matrices in the sense that the respective barycenters (with the same weights) in the uplink and downlink spaces are corresponding.
Specifically, we choose the weights such that $\R^{UL}$ is as close as possible to being a weighted barycenter of the uplink covariance matrices.
In the ideal case where $\R^{UL}$ is the exact barycenter of $(\R_1^{UL},...,\R_{K}^{UL})$ with weights $w_1,...,w_{K}$, this consists in choosing the weights such that
\begin{equation}
\sum_{k=1}^{K} w_k\exp^{-1}_{\R^{UL}}(\R_k^{UL})=0. \label{weights_bary_K}
\end{equation}
In order to avoid ambiguity in this definition (there might be multiple sets of weights fulfilling this property), we restrict our consideration to a set of $K_s\leq K$ uplink covariances closest to $\R^{UL}$ (see below for the choice of $K_s$); we denote this set (up to a change of indices) by $(\R_1^{UL},...,\R_{K_s}^{UL})$. We then choose the interpolation weights as (compare with \eqref{weights_bary_K}):
\begin{equation}
\left\{\begin{array}{ll}
(w_1,...,w_{K_s}) \\
\displaystyle\ \ \ \ \ = \arg\min_{\substack{w_1,...,w_{K_s}\in[0;1]\\ \sum_{k=1}^{K_s} w_k=1}} \left\| \sum_{k=1}^{K_s} w_k\exp^{-1}_{\R^{UL}}(\R_k^{UL})\right\|_F,\\
w_k=0\text{      for $k>K_s$.}
\end{array}\right.
\label{eqself}
\end{equation}
Let us denote ${\bf w}=(w_1,...,w_{K_s})$, and let the matrix $\M$ of size $N_R^2\times K_s$ be the concatenation of the vectorization of the matrices $\exp^{-1}_{\R^{UL}}(\R_k^{UL})$, i.e.
\begin{equation*}
\M=\left(\mathrm{vec}(\exp^{-1}_{\R^{UL}}(\R_1^{UL})), \dots, \mathrm{vec}(\exp^{-1}_{\R^{UL}}(\R_k^{UL}))\right).
\end{equation*}
The minimization problem of eq.~(\ref{eqself}) can be rewritten
\[
(w_1,...,w_{K_s}) = \arg\min_{\substack{{\bf w}=w_1,...,w_{K_s}\in[0;1]\\ \sum_{k=1}^{K_s} w_k=1}} {\bf w}^H\M^H\M {\bf w}.
\]
If $K_s>N_R^2$, then $\text{rank}(\M^H\M)<K_s$, i.e. the nullspace of $\M^H\M$ is nonempty, and the choice of ${\bf w}$ is not unique, which can deteriorate the performances of the interpolation; in order to avoid this under-learning effect, we choose $K_s=\min(N_R^2,K)$.\\

\item \textbf{Kernel interpolation method}\\
Let $\varphi$ be a decreasing function, also referred to as the \emph{kernel function}. We define the barycenter weights based on the distance to the interpolated point:
\begin{equation}
w_i = \frac{\varphi(d(\R_i^{UL},\R^{UL}))} {\sum_{j=1}^K \varphi(d(\R_j^{UL},\R^{UL}))}.
\label{eq_wkernel}
\end{equation}
The barycenter given by eq.~(\ref{eq_barycenter}) with weights given by eq.~(\ref{eq_wkernel}) is the Riemannian equivalent of classical kernel smoothing and more specifically of \emph{kernel regression} \cite{watson64}. Indeed, in the Euclidean space, kernel regression aims to find a non-linear relation between two random variables $X$ and $Y$ by estimating $\mathbb{E}[Y|X]$. In the kernel framework, the density of the couple $(X,Y)$ is estimated from a sample $(x_1,y_1),\dots,(x_K,y_K)$ by
\[
\hat{f}(x,y) = \frac{1}{K}\sum_{k=1}^K f_1\left(x-x_k\right)f_2\left(y-y_k\right),
\]
with $f_1$ and $f_2$ two density functions symmetric with respect to 0. If we observe $x$, the corresponding $y$ is ``regressed'' by the conditional expectation 
\[
\hat{y} =  \sum_{k=1}^K \frac{f_1\left(x-x_k\right)}{\sum_{j=1}^Nf_1\left(x-x_j\right)}y_k.
\]
Using the analogy $x\leftrightarrow \R^{UL}$, $y\leftrightarrow \R$ and $f_1\leftrightarrow \varphi(\|.\|)$, \eqref{eq_barycenter} can be interpreted as an Euclidean barycenter with kernel weights corresponding to eq. (\ref{eq_wkernel}).
Here we propose to use the Gaussian kernel function $\varphi_{\sigma}(r) = \mathrm{e}^{-r^2/2\sigma^2}$. The parameter $\sigma$ (called bandwidth) represents the disparity of the dictionary and must be estimated. We suggest a criterion inspired from the mirror-interpolation method, whereby we minimize the same function as in \eqref{eqself} over the single variable $\sigma$:
\begin{eqnarray}
\nonumber
&&\hspace{-2.5cm}\sigma\left(\R^{UL}\right) = \arg\min_{\sigma>0}\\
&&\hspace{-1.5cm} \left\| \sum_{k=1}^K \frac{\varphi_{\sigma}(d(\R_k^{UL},\R^{UL}))} {\sum_{k=1}^K\varphi_{\sigma}(d(\R_k^{UL},\R^{UL}))} \exp^{-1}_{\R^{UL}}(\R_k^{UL})\right\|_F.
\label{eq_bandwidth}
\end{eqnarray}
\end{itemize}


\section{Performance Evaluation}
\label{section_performance}

We now present simulated performance results corresponding to the various interpolation methods presented in the previous section, and compare them with state-of-the-art techniques.

\subsection{Simulated scenario}

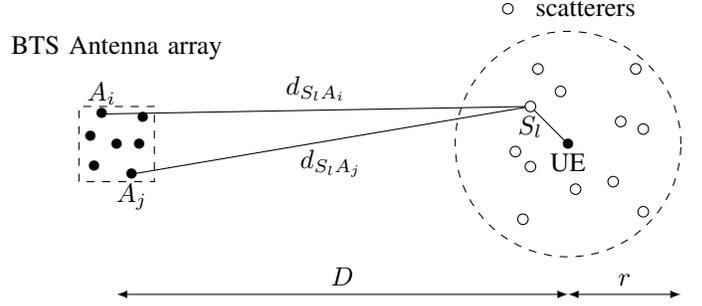
\begin{figure}
\centering
\begin{tikzpicture}
\coordinate (A) at (-0.2,0.4);
\coordinate (A2) at (0.2,-0.4);
\coordinate (S) at (5.5,0.5);
\coordinate (center) at (6,0);
\draw [dashed] (-0.5,-0.5) rectangle (0.5,0.5);
\draw (A) node {$\bullet$};
\draw (A) node[above] {$A_i$};
\draw (A2) node {$\bullet$};
\draw (A2) node[below] {$A_j$};
\draw (-0.3,-0.3) node {$\bullet$};
\draw (0.35,0.35) node {$\bullet$};
\draw (-0.35,0.1) node {$\bullet$};
\draw (0,0) node {$\bullet$};
\draw (0.3,0) node {$\bullet$};
\draw (A2) -- (S) node[midway,below] {$d_{S_lA_j}$};
\draw (A) -- (S) node[midway,above] {$d_{S_lA_i}$};
\draw (S) node[below] {$S_l$};
\tikzstyle{every node}=[draw,circle,fill=white,minimum size=4pt,
                            inner sep=0pt]
\draw (S) -- (center);
\draw (S) node {};
\draw (5.6,1) node {};
\draw (6.9,1) node {};
\draw (6.7,0.3) node {};
\draw (5.4,-1.0) node {};
\draw (6.6,-0.5) node {};
\draw (5.9,0.7) node {};
\draw (7,-0.9) node {};
\draw (5.5,-0.3) node {};
\draw (5.3,-0.1) node {};
\draw (6.1,-0.6) node {};
\draw (7.0,0.2) node {};
\draw (5.2,1.8) node {};
  \tikzstyle{every node}=[];
\draw (5.2,1.8) node[right] {\ \ scatterers};
\draw [dashed] (center) circle (1.5);
\draw (center) node {$\bullet$};
\draw (center) node[below] {UE};
\draw (0,1) node[above] {BTS Antenna array};
\draw[<->,>=latex] (0,-2) -- (6,-2) node[midway,above] {$D$};
\draw[<->,>=latex] (6,-2) -- (7.5,-2) node[midway,above] {$r$};
\end{tikzpicture}
\caption{Illustration of a ray tracing model: the UE antenna is assumed to be surrounded by scatterers randomly located in a ball of radius $r$ (black and white bullets respectively represent antennas and scatterers).}
\label{figmodel}
\end{figure}

We consider an array of $N_R=N_T$ antennas communicating with a single user at distance $D$ uniformly distributed in the interval $[100,900]$ meters. In order to capture the limited angular spread under which the UE is seen at the BTS, the channel is generated using the ring model of \cite{shiu00}, with scatterers uniformly distributed in a ball of radius $r\in [1,100]$ meters containing $N_S=1000$ scatter points (see Figure \ref{figmodel}).\\
A ray tracing model with a simple quadratic pathloss is assumed, whereby the covariance $\R=(R_{ij})_{1\leq i,j\leq N}$ of the channel at wavelength $\lambda$ is then modeled by \cite{shiu00}
\begin{equation}
R_{ij} = \frac{P}{D^2N_S}\sum_{l=1}^{N_S} e^{2i\frac{\pi}{\lambda}(d_{S_lA_i}-d_{S_lA_j})} +P_N \delta_{ij},
\label{eq_cov_model}
\end{equation}
where $P$ is the received power at the user side and $P_N$ the power of thermal noise, $d_{S_lA_i}$ denotes the distance between the $l$-th scatterer and the $i$-th BTS antenna, and with $\delta_{ii}=1$ and $\delta_{ij}=0$ if $i\neq j$.
%
%

The parameters of the simulated scenario are the following: 
$K$ (equal to $50,100,150,300$ or $500$) pairs of uplink/dowlink correlation matrices in the dictionary. Each pair is generated by randomly placing a single-antenna user surrounded by a ball of scatterers (Fig. \ref{figmodel}). The random position of the UE antenna is taken uniformly in the area at distance comprised around $100$m to $900$m around the antenna array. The $N_R=N_T=10$ BTS antennas are placed either in a uniform linear array (denoted by ULA), or randomly drawn uniformly in a square (denoted by random array geometry). The uplink and dowlink channels operate at different frequencies (downlink at $1.8$ GHz, uplink at $1.9$ GHz or $2.8$ GHz).
For each position of the UE, the uplink/downlink covariance matrices are obtained through the sample covariance computed from $L=1000$ realizations of the channel $\h$. A realization of the channel is the random vector $\h=\R^{\frac{1}{2}}\w$ with $\w\sim \mathcal{CN}(0,I_N)$. Note that $\R$ depends on the UE location according to \eqref{eq_cov_model}.

We will compare the three dictionary-based estimation procedures of Section~\ref{section_interpolation} (namely, nearest neighbor, mirror interpolation, and kernel interpolation) with the following approaches:
\begin{itemize}
\item \textbf{No conversion}: the downlink covariance matrix is estimated by assuming that it is equal to the uplink covariance matrix: $\hat{\R}=\R^{UL}$ (note that this is possible since we assume $N_R=N_T$ in our simulation setup).
\item \textbf{Spline-based interpolation}: this is the geometry-based interpolation scheme from \cite{jordan09}, which relies on the fact that the downlink covariance function is a dilatation of the uplink covariance function in the case of a ULA. The covariance coefficients are interpolated by a cubic spline. 
\item \textbf{Perfect feedback}: we assume that the downlink covariance is estimated by the UE and fed back without quantization to the BTS. This constitutes a best-case performance benchmark, and neglects the cost of over-the-air feedback. $\hat{\R}$ remains affected by the estimation noise due to the sample covariance estimator.
\end{itemize}

\subsection{Performance Results}
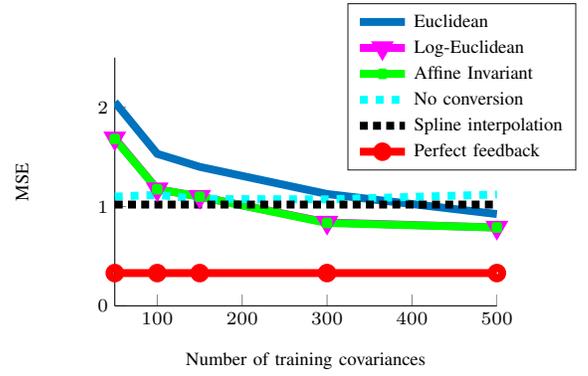
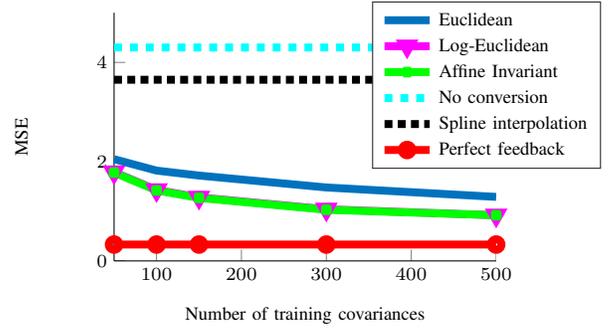
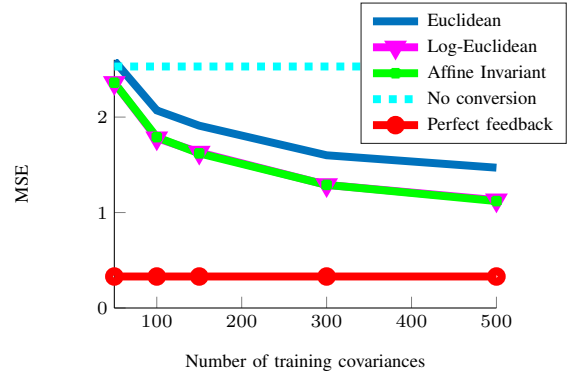
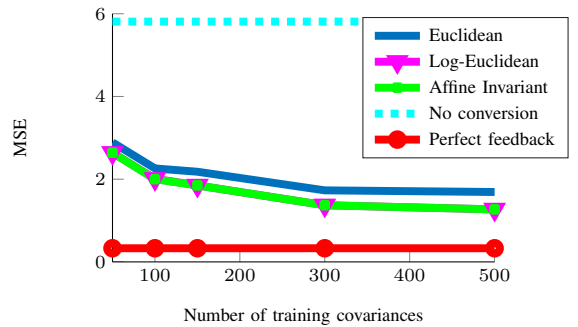
\begin{figure}
	\centering
   \subfloat[Uplink and downlink frequencies respectively equal to 1.8GHz and 1.9GHz (ULA).] {\input{mse_errors_nearest_1819.tikz}}\\
   \subfloat[Uplink and downlink frequencies respectively equal to 1.8GHz and 2.8GHz (ULA).] {\input{mse_errors_nearest_1828.tikz}}\\

   \subfloat[Uplink and downlink frequencies respectively equal to 1.8GHz and 1.9GHz (Random array geometry).] {\input{mse_errors_nearest_rgeom_1819.tikz}}\\
   \subfloat[Uplink and downlink frequencies respectively equal to 1.8GHz and 2.8GHz (Random array geometry).] {\input{mse_errors_nearest_rgeom_1828.tikz}}
	\caption{Mean Square Error vs. dictionary size for the estimation of the downlink covariance matrix for the nearest neighbor method.}
	\label{fig_mse}
\end{figure}

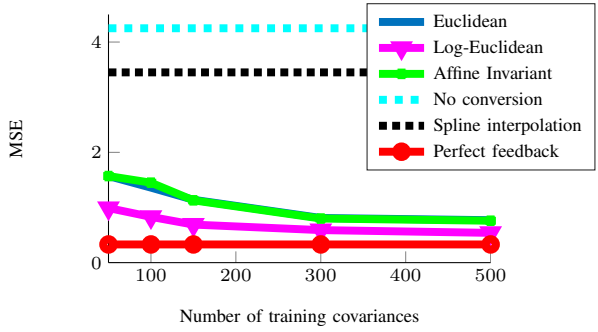
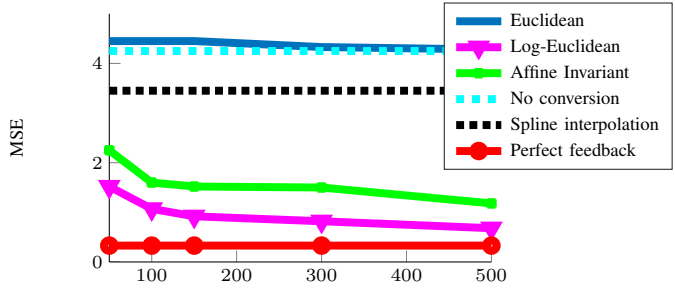
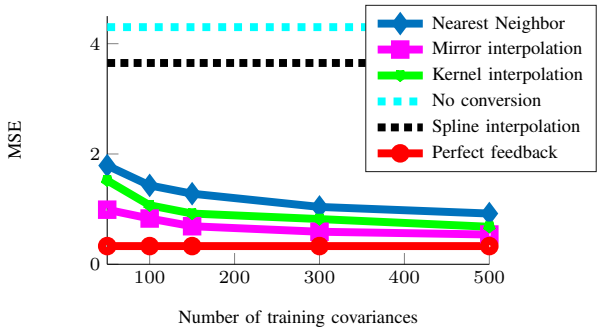
\begin{figure}
	\centering
   \subfloat[Mirror-interpolation method.] {\input{mse_errors_autointerp_pos_1828.tikz}}\\
   \subfloat[Kernel interpolation method.] {\input{mse_errors_kernel_1828.tikz}}\\
   \subfloat[Comparison of interpolation methods for the log-Euclidean metric.] {\input{mse_errors_le_pos_1828.tikz}}
	\caption{Mean Square Error vs. dictionary size for the estimation of the downlink covariance matrix with uplink and downlink frequencies respectively equal to 1.8GHz and 2.8GHz (ULA).}
	\label{fig_mse2}
\end{figure}

We evaluate the performance of the proposed method in terms of the quality of the estimation of the downlink covariance matrix itself, through the average mean square error $d_{AI}(\R,\hat{\R})^2$ obtained by Monte-Carlo simulation.
The results are presented in Figures \ref{fig_mse} and \ref{fig_mse2} for two possible choices of the frequency gap between the uplink and downlink bands ($1.8/1.9$GHz and $1.8/2.8$GHz). Note that the spline-based method and the perfect feedback method are not dictionary-based, and therefore their performance does not depend on $K$.

Figure~\ref{fig_mse} presents a comparison between the nearest-neighbor dictionary based method (evaluated for various choices of the distance on $\mathcal{S}_N^{++}(\mathbb{C})$) with three reference approaches (no conversion, spline approximation, and perfect feedback).
It can be observed that the Nearest Neighbor estimator performs better than the spline method and the naive method consisting in assuming that $\hat{\R}=\R^{UL}$, when the frequency gap is big or the geometry of antenna random. Indeed, the dilatation hypothesis of the spline method requires the frequency gap to be small. Moreover, when the geometry of antennas is random, the spline method is unappplicable whereas the performances of interpolation-based estimators are similar to the uniform linear array case.

In Figure~\ref{fig_mse2}, we compare the performance of the various dictionary-based methods proposed in this article (mirror interpolation, kernel interpolation, nearest neighbor).
The gains of the kernel and the mirror-interpolation with respect to the nearest neighbor method for non-Euclidean geometry can be explained by the exploitation of the information of multiple neighbors instead of one. The log-Euclidean metric coupled with the kernel interpolation or the mirror-interpolation provides the optimal performance among the considered methods. As for the choice of the metric, the log-Euclidean and the affine invariant ones outperform the Euclidean metric that does not take into account the constraints of the space of positive definite matrices.


\section{Complexity of interpolation schemes}
\label{sec_complexity}
The complexity of the interpolation scheme and of the computation of the weights is summarized in Table~\ref{table_complexity}. $C(K,N_T)$ represents the complexity of the Riemannian gradient descent used for the computation of the barycenter for the affine invariant metric; since this is an iterative algorithm, this complexity is difficult to evaluate as it depends on the required accuracy.
The complexity of the interpolation methods is strongly related to the complexity of the computation of $K$ logarithms of matrices of size $N\times N$ (equal to $O(KN^3)$ for a simple algorithm). In practice, the gradient descent is the limiting factor for the affine invariant metric.
The log-Euclidean metric acts like Euclidean metric once we consider the logarithm of every covariance matrix. The complexity then differs only from the computation of these matrix logarithms.
\begin{table}
\footnotesize
\centering
\begin{tabular}{|c|c|c|c|}
\hline
&Euclidean &log-Euclidean & affine invariant\\
\hline
Nearest Neighbor & $KN^2$ & $KN^3$ & $KN^3$\\
Mirror-interpolation & $KN^4$ & $KN^4$ & $KN^3C(K,N_T)$\\
Kernel method & $KN^3$ & $KN^3$ & $KN^3 C(K,N_T)$\\
\hline
\end{tabular}
\caption{Complexity order of the proposed interpolation schemes (with $N=N_T+N_R$).}
\label{table_complexity}
\end{table}
\begin{table}
\footnotesize
\centering
\begin{tabular}{|c|c|c|c|}
\hline
&Euclidean &log-Euclidean & affine invariant\\
\hline
Nearest Neighbor & 2.1 & 32.5 & 185 \\
Mirror-interpolation & 3.1 & 52.7 & 206 \\
Kernel method & 5 & 56.5 & 240 \\
\hline
\end{tabular}
\caption{Computation time (in milliseconds) of interpolation schemes ($K=50$, $N_T=N_R=10$).}
\label{table_time}
\end{table}
Table~\ref{table_time} shows the elapsed time for the computation of one interpolation matrix for a DualCore processor (at $1.9$GHz/$2.5$GHz). The log-Euclidean metric provides a favorable efficiency/speed trade-off.


\section{Conclusion}
We presented estimators of the covariance of the downlink channel between a base station and a single-antenna UE interpolated from the uplink covariance matrix and a dictionary of uplink/dowlink pairs of covariance matrices collected during a training phase. The simulations illustrate the advantages of the interpolation schemes in the case where the array geometry is random and the carrier frequency gap is large. In particular, amongst the considered methods, the kernel interpolation and mirror-interpolation used with a log-Euclidean metric emerge as estimators combining performance and reasonably low computational requirements. 


\bibliographystyle{IEEEtran}

\bibliography{dummy_biblio,maxime_biblio}
%
%
%

\end{document}

%% file: learning_phase.tikz
%
%
\definecolor{mycolor1}{rgb}{0.00000,0.44700,0.74100}%
\definecolor{mycolor2}{rgb}{1.00000,0.00000,1.00000}%
\definecolor{mycolor3}{rgb}{0.00000,1.00000,1.00000}%

%
\begin{tikzpicture}
\scriptsize
\coordinate (A1) at (-0.5,0);
\coordinate (A2) at (0.25,-0.4);
\coordinate (A3) at (0.25,0.4);
\coordinate (A4) at (0.5,0);
\coordinate (A5) at (-0.25,0.4);
\coordinate (A6) at (-0.25,-0.4);
\coordinate (UE1) at (3,0.5);
\coordinate (UE3) at (2.5,-1);
\coordinate (UE4) at (0.5,-2);
\coordinate (UE5) at (0,2);
\coordinate (UE6) at (-2,0.5);

\draw (A1) -- ++(0,-0.2);
\draw (A1) -- ++(0.1,0.05);
\draw (A1) -- ++(-0.1,0.05);
\draw (A2) -- ++(0,-0.2);
\draw (A2) -- ++(0.1,0.05);
\draw (A2) -- ++(-0.1,0.05);
\draw (A3) -- ++(0,-0.2);
\draw (A3) -- ++(0.1,0.05);
\draw (A3) -- ++(-0.1,0.05);
\draw (A4) -- ++(0,-0.2);
\draw (A4) -- ++(0.1,0.05);
\draw (A4) -- ++(-0.1,0.05);
\draw (A5) -- ++(0,-0.2);
\draw (A5) -- ++(0.1,0.05);
\draw (A5) -- ++(-0.1,0.05);
\draw (A6) -- ++(0,-0.2);
\draw (A6) -- ++(0.1,0.05);
\draw (A6) -- ++(-0.1,0.05);
\draw (0,-0.05) node[align=center,text width=1cm] {Base station};

\draw[line width=1pt] (UE1) circle (0.4) node [below=0.5cm] {UE at position 1};
\draw[line width=1pt] (UE3) circle (0.5) node [below=0.5cm] {UE at position 2};
\draw[line width=1pt] (UE4) circle (0.5) node [below=0.5cm] {UE at position 3};
\draw[line width=1pt] (UE5) circle (0.5) node [above=0.5cm] {UE at position 4};
\draw[line width=1pt] (UE6) circle (0.5) node [above=0.5cm] {UE at position 5};

\draw[->,>=latex] (0.7,0.1) -- (2.3,0.5) node[midway,above] {$R_1$};
\draw[<-,>=latex] (0.7,0) -- (2.3,0.4) node[midway,below] {$R_1^{UL}$};
\draw[<-,>=latex] (0.7,-0.4) -- (2,-0.8) node[midway,above=0.05cm] {$R_2$};
\draw[->,>=latex] (0.7,-0.3) -- (2,-0.7) node[midway,below=0.1cm] {$R_2^{UL}$};
\draw[<-,>=latex] (0.3,-0.7) -- (0.4,-1.4) node[midway,right=0.05cm] {$R_3$};
\draw[->,>=latex] (0.4,-0.7) -- (0.5,-1.4) node[midway,left=0.05cm] {$R_3^{UL}$};
\draw[<-,>=latex] (0,0.7) -- (0,1.4) node[midway,right=0.05cm] {$R_4$};
\draw[->,>=latex] (0.1,0.7) -- (0.1,1.4) node[midway,left=0.05cm] {$R_4^{UL}$};
\draw[->,>=latex] (-0.7,0.2) -- (-1.3,0.4) node[midway,above=0.05cm] {$R_5$};
\draw[<-,>=latex] (-0.7,0.1) -- (-1.3,0.3) node[midway,below=0.05cm] {$R_5^{UL}$};

\draw[line width=1pt] (-3.5,-2) circle (0.25);
\draw (-3,-2) node[right,text width=3cm] {User Equipment during training period};

\end{tikzpicture}%

%% file: exploitation_phase.tikz
%
%
\definecolor{mycolor1}{rgb}{0.00000,0.44700,0.74100}%
\definecolor{mycolor2}{rgb}{1.00000,0.00000,1.00000}%
\definecolor{mycolor3}{rgb}{0.00000,1.00000,1.00000}%

%
\begin{tikzpicture}
\scriptsize
\coordinate (A1) at (-0.5,0);
\coordinate (A2) at (0.25,-0.4);
\coordinate (A3) at (0.25,0.4);
\coordinate (A4) at (0.5,0);
\coordinate (A5) at (-0.25,0.4);
\coordinate (A6) at (-0.25,-0.4);
\coordinate (UE1) at (3,0.2);
\coordinate (UE3) at (2.5,-1.3);
\coordinate (UE4) at (0.5,-2);
\coordinate (UE5) at (0,2);
\coordinate (UE6) at (-2,0.5);
\coordinate (UE_test) at (2.2,2);

\draw (A1) -- ++(0,-0.2);
\draw (A1) -- ++(0.1,0.05);
\draw (A1) -- ++(-0.1,0.05);
\draw (A2) -- ++(0,-0.2);
\draw (A2) -- ++(0.1,0.05);
\draw (A2) -- ++(-0.1,0.05);
\draw (A3) -- ++(0,-0.2);
\draw (A3) -- ++(0.1,0.05);
\draw (A3) -- ++(-0.1,0.05);
\draw (A4) -- ++(0,-0.2);
\draw (A4) -- ++(0.1,0.05);
\draw (A4) -- ++(-0.1,0.05);
\draw (A5) -- ++(0,-0.2);
\draw (A5) -- ++(0.1,0.05);
\draw (A5) -- ++(-0.1,0.05);
\draw (A6) -- ++(0,-0.2);
\draw (A6) -- ++(0.1,0.05);
\draw (A6) -- ++(-0.1,0.05);
\draw (0,-0.05) node[align=center,text width=1cm] {Base station};

\draw[line width=1pt] (UE1) circle (0.4) node [below=0.5cm] {UE at position 1};
\draw[line width=1pt] (UE3) circle (0.5) node [below=0.5cm] {UE at position 2};
\draw[line width=1pt] (UE4) circle (0.5) node [below=0.5cm] {UE at position 3};
\draw[line width=1pt] (UE5) circle (0.5) node [above=0.5cm] {UE at position 4};
\draw[line width=1pt] (UE6) circle (0.5) node [above=0.5cm] {UE at position 5};
\draw[line width=1pt,dashed] (UE_test) circle (0.5) node [above=0.5cm] {UE at new position};

\draw[->,>=latex] (0.7,0.1) -- (2.4,0.2) node[midway,above] {$R_1$};
\draw[<-,>=latex] (0.7,0) -- (2.4,0.1) node[midway,below] {$R_1^{UL}$};
\draw[<-,>=latex] (0.7,-0.4) -- (2,-1.1) node[midway,above=0.05cm] {$R_2$};
\draw[->,>=latex] (0.7,-0.3) -- (2,-1) node[midway,below=0.15cm] {$R_2^{UL}$};
\draw[<-,>=latex] (0.3,-0.7) -- (0.4,-1.4) node[midway,right=0.05cm] {$R_3$};
\draw[->,>=latex] (0.4,-0.7) -- (0.5,-1.4) node[midway,left=0.05cm] {$R_3^{UL}$};
\draw[<-,>=latex] (0,0.7) -- (0,1.4) node[midway,right=0.05cm] {$R_4$};
\draw[->,>=latex] (0.1,0.7) -- (0.1,1.4) node[midway,left=0.05cm] {$R_4^{UL}$};
\draw[->,>=latex] (-0.7,0.2) -- (-1.3,0.4) node[midway,above=0.05cm] {$R_5$};
\draw[<-,>=latex] (-0.7,0.1) -- (-1.3,0.3) node[midway,below=0.05cm] {$R_5^{UL}$};
\draw[<-,>=latex] (0.5,0.5) -- (1.7,1.6) node[midway,above=0.1cm] {$\bf R^{UL}$};
\draw[->,>=latex] (0.5,0.4) -- (1.7,1.5) node[midway,right=0.2cm] {\bf Corresponding $\bf R$?};

\draw[line width=1pt] (-3.8,-2) circle (0.25);
\draw (-3.3,-2) node[right,text width=3cm] {User Equipment during training period};
\draw[line width=1pt,dashed] (-3.8,-2.8) circle (0.25);
\draw (-3.3,-2.8) node[right,text width=3cm] {User Equipment during exploitation period};

\end{tikzpicture}%

%% file: mse_errors_nearest_1819.tikz
%
%
\definecolor{mycolor1}{rgb}{0.00000,0.44700,0.74100}%
\definecolor{mycolor2}{rgb}{1.00000,0.00000,1.00000}%
\definecolor{mycolor3}{rgb}{0.00000,1.00000,1.00000}%
\begin{tikzpicture}
\scriptsize

\begin{axis}[%
width=2in,
height=1.3in,
at={(1.558646in,0.7887in)},
scale only axis,
xmin=50,
xmax=500,
xlabel={Number of training covariances},
ymin=0,
ymax=2.5,
ylabel={MSE},
axis x line*=bottom,
axis y line*=left,
legend style={at={(0.608722,0.545269)},anchor=south west,legend cell align=left,align=left,draw=white!15!black}
]
\addplot [color=mycolor1,solid,line width=3.0pt]
  table[row sep=crcr]{%
50	2.05285936555773\\
100	1.53116371881946\\
150	1.4\\
300	1.12648408230061\\
500	0.924502562830786\\
};
\addlegendentry{Euclidean};

\addplot [color=mycolor2,solid,line width=3.0pt,mark=triangle,mark options={solid,rotate=180}]
  table[row sep=crcr]{%
50	1.6882280782095\\
100	1.17585339338239\\
150	1.1\\
300	0.83984987437714\\
500	0.788994442074001\\
};
\addlegendentry{Log-Euclidean};

\addplot [color=green,solid,line width=3.0pt,mark=+,mark options={solid}]
  table[row sep=crcr]{%
50	1.6800348451609\\
100	1.17427867876348\\
150	1.1\\
300	0.834701863731\\
500	0.789000594507484\\
};
\addlegendentry{Affine Invariant};

\addplot [color=mycolor3,dashed,line width=3.0pt]
  table[row sep=crcr]{%
50	1.1\\
100	1.11804089647524\\
150	1.07600676690291\\
300	1.07269919667355\\
500	1.12179817285894\\
};
\addlegendentry{No conversion};

\addplot [color=black,dotted,line width=3.0pt]
  table[row sep=crcr]{%
50	1.02\\
100	1.02\\
150	1.02\\
300	1.02\\
500	1.02\\
};
\addlegendentry{Spline interpolation};

\addplot [color=red,solid,line width=3.0pt,mark=o,mark options={solid}]
  table[row sep=crcr]{%
50	0.33\\
100	0.33\\
150	0.33\\
300	0.33\\
500	0.33\\
};
\addlegendentry{Perfect feedback};

\end{axis}
\end{tikzpicture}%

%% file: mse_errors_nearest_1828.tikz
%
%
\definecolor{mycolor1}{rgb}{0.00000,0.44700,0.74100}%
\definecolor{mycolor2}{rgb}{1.00000,0.00000,1.00000}%
\definecolor{mycolor3}{rgb}{0.00000,1.00000,1.00000}%
\begin{tikzpicture}
\scriptsize

\begin{axis}[%
width=2in,
height=1.3in,
at={(1.558646in,0.7887in)},
scale only axis,
xmin=50,
xmax=500,
xlabel={Number of training covariances},
ymin=0,
ymax=5,
ylabel={MSE},
axis x line*=bottom,
axis y line*=left,
legend style={at={(0.675291,0.37128)},anchor=south west,legend cell align=left,align=left,draw=white!15!black}
]
\addplot [color=mycolor1,solid,line width=3.0pt]
  table[row sep=crcr]{%
50	2.05\\
100	1.82\\
150	1.72\\
300	1.48\\
500	1.29\\
};
\addlegendentry{Euclidean};

\addplot [color=mycolor2,solid,line width=3.0pt,mark=triangle,mark options={solid,rotate=180}]
  table[row sep=crcr]{%
50	1.79\\
100	1.43\\
150	1.28\\
300	1.04\\
500	0.92\\
};
\addlegendentry{Log-Euclidean};

\addplot [color=green,solid,line width=3.0pt,mark=+,mark options={solid}]
  table[row sep=crcr]{%
50	1.78\\
100	1.42\\
150	1.27\\
300	1.03\\
500	0.92\\
};
\addlegendentry{Affine Invariant};

\addplot [color=mycolor3,dashed,line width=3.0pt]
  table[row sep=crcr]{%
50	4.3\\
100	4.3\\
150	4.3\\
300	4.3\\
500	4.3\\
};
\addlegendentry{No conversion};

\addplot [color=black,dotted,line width=3.0pt]
  table[row sep=crcr]{%
50	3.65\\
100	3.65\\
150	3.65\\
300	3.65\\
500	3.65\\
};
\addlegendentry{Spline interpolation};

\addplot [color=red,solid,line width=3.0pt,mark=o,mark options={solid}]
  table[row sep=crcr]{%
50	0.33\\
100	0.33\\
150	0.33\\
300	0.33\\
500	0.33\\
};
\addlegendentry{Perfect feedback};

\end{axis}
\end{tikzpicture}%


%% file: mse_errors_nearest_rgeom_1819.tikz
%
%
\definecolor{mycolor1}{rgb}{0.00000,0.44700,0.74100}%
\definecolor{mycolor2}{rgb}{1.00000,0.00000,1.00000}%
\definecolor{mycolor3}{rgb}{0.00000,1.00000,1.00000}%
\begin{tikzpicture}
\scriptsize

\begin{axis}[%
width=2in,
height=1.3in,
at={(1.558646in,0.671458in)},
scale only axis,
xmin=50,
xmax=500,
xlabel={Number of training covariances},
ymin=0,
ymax=2.6,
ylabel={MSE},
axis x line*=bottom,
axis y line*=left,
legend style={at={(0.644342,0.661116)},anchor=south west,legend cell align=left,align=left,draw=white!15!black}
]
\addplot [color=mycolor1,solid,line width=3.0pt]
  table[row sep=crcr]{%
50	2.59\\
100	2.07\\
150	1.91\\
300	1.60\\
500	1.47\\
};
\addlegendentry{Euclidean};

\addplot [color=mycolor2,solid,line width=3.0pt,mark=triangle,mark options={solid,rotate=180}]
  table[row sep=crcr]{%
50	2.36\\
100	1.78\\
150	1.63\\
300	1.29\\
500	1.13\\
};
\addlegendentry{Log-Euclidean};

\addplot [color=green,solid,line width=3.0pt,mark=+,mark options={solid}]
  table[row sep=crcr]{%
50	2.36\\
100	1.79\\
150	1.62\\
300	1.29\\
500	1.12\\
};
\addlegendentry{Affine Invariant};

\addplot [color=mycolor3,dashed,line width=3.0pt]
  table[row sep=crcr]{%
50	2.53\\
100	2.53\\
150	2.53\\
300	2.53\\
500	2.53\\
};
\addlegendentry{No conversion};

\addplot [color=red,solid,line width=3.0pt,mark=o,mark options={solid}]
  table[row sep=crcr]{%
50	0.33\\
100	0.33\\
150	0.33\\
300	0.33\\
500	0.33\\
};
\addlegendentry{Perfect feedback};

\end{axis}
\end{tikzpicture}%


%% file: mse_errors_nearest_rgeom_1828.tikz
%
%
\definecolor{mycolor1}{rgb}{0.00000,0.44700,0.74100}%
\definecolor{mycolor2}{rgb}{1.00000,0.00000,1.00000}%
\definecolor{mycolor3}{rgb}{0.00000,1.00000,1.00000}%
\begin{tikzpicture}
\scriptsize

\begin{axis}[%
width=2in,
height=1.3in,
at={(1.558646in,0.671458in)},
scale only axis,
xmin=50,
xmax=500,
xlabel={Number of training covariances},
ymin=0,
ymax=6,
ylabel={MSE},
axis x line*=bottom,
axis y line*=left,
legend style={at={(0.653936,0.417182)},anchor=south west,legend cell align=left,align=left,draw=white!15!black}
]
\addplot [color=mycolor1,solid,line width=3.0pt]
  table[row sep=crcr]{%
50	2.89\\
100	2.26\\
150	2.18\\
300	1.73\\
500	1.69\\
};
\addlegendentry{Euclidean};

\addplot [color=mycolor2,solid,line width=3.0pt,mark=triangle,mark options={solid,rotate=180}]
  table[row sep=crcr]{%
50	2.64\\
100	2.01\\
150	1.85\\
300	1.37\\
500	1.27\\
};
\addlegendentry{Log-Euclidean};

\addplot [color=green,solid,line width=3.0pt,mark=+,mark options={solid}]
  table[row sep=crcr]{%
50	2.64\\
100	2.00\\
150	1.85\\
300	1.37\\
500	1.27\\
};
\addlegendentry{Affine Invariant};

\addplot [color=mycolor3,dashed,line width=3.0pt]
  table[row sep=crcr]{%
50	5.81\\
100	5.81\\
150	5.81\\
300	5.81\\
500	5.81\\
};
\addlegendentry{No conversion};

\addplot [color=red,solid,line width=3.0pt,mark=o,mark options={solid}]
  table[row sep=crcr]{%
50	0.33\\
100	0.33\\
150	0.33\\
300	0.33\\
500	0.33\\
};
\addlegendentry{Perfect feedback};

\end{axis}
\end{tikzpicture}%


%% file: mse_errors_autointerp_pos_1828.tikz
%
%
\definecolor{mycolor1}{rgb}{0.00000,0.44700,0.74100}%
\definecolor{mycolor2}{rgb}{1.00000,0.00000,1.00000}%
\definecolor{mycolor3}{rgb}{0.00000,1.00000,1.00000}%
\begin{tikzpicture}
\scriptsize

\begin{axis}[%
width=2in,
height=1.3in,
at={(1.558646in,0.7887in)},
scale only axis,
xmin=50,
xmax=500,
xlabel={Number of training covariances},
ymin=0,
ymax=4.5,
ylabel={MSE},
axis x line*=bottom,
axis y line*=left,
legend style={at={(0.675291,0.37128)},anchor=south west,legend cell align=left,align=left,draw=white!15!black}
]
\addplot [color=mycolor1,solid,line width=3.0pt]
  table[row sep=crcr]{%
50	1.57\\
100	1.36\\
150	1.14\\
300	0.81\\
500	0.77\\
};
\addlegendentry{Euclidean};

\addplot [color=mycolor2,solid,line width=3.0pt,mark=triangle,mark options={solid,rotate=180}]
  table[row sep=crcr]{%
50	0.99\\
100	0.83\\
150	0.69\\
300	0.59\\
500	0.54\\
};
\addlegendentry{Log-Euclidean};

\addplot [color=green,solid,line width=3.0pt,mark=+,mark options={solid}]
  table[row sep=crcr]{%
50	1.57\\
100  1.45\\
150	1.13\\
300	0.80\\
500	 0.76\\
};
\addlegendentry{Affine Invariant};

\addplot [color=mycolor3,dashed,line width=3.0pt]
  table[row sep=crcr]{%
50	4.25\\
100	4.25\\
150	4.25\\
300	4.25\\
500	4.25\\
};
\addlegendentry{No conversion};

\addplot [color=black,dotted,line width=3.0pt]
  table[row sep=crcr]{%
50	3.45\\
100	3.45\\
150	3.45\\
300	3.45\\
500	3.45\\
};
\addlegendentry{Spline interpolation};

\addplot [color=red,solid,line width=3.0pt,mark=o,mark options={solid}]
  table[row sep=crcr]{%
50	0.33\\
100	0.33\\
150	0.33\\
300	0.33\\
500	0.33\\
};
\addlegendentry{Perfect feedback};

\end{axis}
\end{tikzpicture}%


%% file: mse_errors_kernel_1828.tikz
%
%
\definecolor{mycolor1}{rgb}{0.00000,0.44700,0.74100}%
\definecolor{mycolor2}{rgb}{1.00000,0.00000,1.00000}%
\definecolor{mycolor3}{rgb}{0.00000,1.00000,1.00000}%
\begin{tikzpicture}
\scriptsize

\begin{axis}[%
width=2in,
height=1.3in,
at={(1.558646in,0.7887in)},
scale only axis,
xmin=50,
xmax=500,
xlabel={Number of training covariances},
ymin=0,
ymax=5,
ylabel={MSE},
axis x line*=bottom,
axis y line*=left,
legend style={at={(0.875291,0.37128)},anchor=south west,legend cell align=left,align=left,draw=white!15!black}
]
\addplot [color=mycolor1,solid,line width=3.0pt]
  table[row sep=crcr]{%
50	4.45\\
100	4.45\\
150	4.45\\
300	4.33\\
500	4.28\\
};
\addlegendentry{Euclidean};

\addplot [color=mycolor2,solid,line width=3.0pt,mark=triangle,mark options={solid,rotate=180}]
  table[row sep=crcr]{%
50	1.52\\
100	1.07\\
150	0.92\\
300	0.82\\
500	0.68\\
};
\addlegendentry{Log-Euclidean};

\addplot [color=green,solid,line width=3.0pt,mark=+,mark options={solid}]
  table[row sep=crcr]{%
50	2.25\\
100 1.60\\
150	1.52\\
300	1.50\\
500	1.18\\
};
\addlegendentry{Affine Invariant};

\addplot [color=mycolor3,dashed,line width=3.0pt]
  table[row sep=crcr]{%
50	4.25\\
100	4.25\\
150	4.25\\
300	4.25\\
500	4.25\\
};
\addlegendentry{No conversion};

\addplot [color=black,dotted,line width=3.0pt]
  table[row sep=crcr]{%
50	3.45\\
100	3.45\\
150	3.45\\
300	3.45\\
500	3.45\\
};
\addlegendentry{Spline interpolation};

\addplot [color=red,solid,line width=3.0pt,mark=o,mark options={solid}]
  table[row sep=crcr]{%
50	0.33\\
100	0.33\\
150	0.33\\
300	0.33\\
500	0.33\\
};
\addlegendentry{Perfect feedback};

\end{axis}
\end{tikzpicture}%


%% file: mse_errors_le_pos_1828.tikz
%
%
\definecolor{mycolor1}{rgb}{0.00000,0.44700,0.74100}%
\definecolor{mycolor2}{rgb}{1.00000,0.00000,1.00000}%
\definecolor{mycolor3}{rgb}{0.00000,1.00000,1.00000}%
\begin{tikzpicture}
\scriptsize

\begin{axis}[%
width=2in,
height=1.3in,
at={(1.558646in,0.7887in)},
scale only axis,
xmin=50,
xmax=500,
xlabel={Number of training covariances},
ymin=0,
ymax=4.5,
ylabel={MSE},
axis x line*=bottom,
axis y line*=left,
legend style={at={(0.675291,0.37128)},anchor=south west,legend cell align=left,align=left,draw=white!15!black}
]
\addplot [color=mycolor1,solid,line width=3.0pt,mark=diamond]
  table[row sep=crcr]{%
50	1.79\\
100	1.43\\
150	1.28\\
300	1.04\\
500	0.92\\
};
\addlegendentry{Nearest Neighbor};

\addplot [color=mycolor2,solid,line width=3.0pt,mark=square,mark options={solid,rotate=180}]
  table[row sep=crcr]{%
50	0.99\\
100	0.83\\
150	0.69\\
300	0.59\\
500	0.54\\
};
\addlegendentry{Mirror interpolation};

\addplot [color=green,solid,line width=3.0pt,mark=star,mark options={solid}]
  table[row sep=crcr]{%
50	1.52\\
100	1.07\\
150	0.92\\
300	0.82\\
500	0.68\\
};
\addlegendentry{Kernel interpolation};

\addplot [color=mycolor3,dashed,line width=3.0pt]
  table[row sep=crcr]{%
50	4.3\\
100	4.3\\
150	4.3\\
300	4.3\\
500	4.3\\
};
\addlegendentry{No conversion};

\addplot [color=black,dotted,line width=3.0pt]
  table[row sep=crcr]{%
50	3.65\\
100	3.65\\
150	3.65\\
300	3.65\\
500	3.65\\
};
\addlegendentry{Spline interpolation};

\addplot [color=red,solid,line width=3.0pt,mark=o,mark options={solid}]
  table[row sep=crcr]{%
50	0.33\\
100	0.33\\
150	0.33\\
300	0.33\\
500	0.33\\
};
\addlegendentry{Perfect feedback};

\end{axis}
\end{tikzpicture}%
